\documentclass[aps,twocolumn,showpacs,preprintnumbers,prb]{revtex4}

\usepackage{graphicx}
\usepackage{dcolumn}
\usepackage{bm}

\newcommand{\beq}{\begin{equation}}
\newcommand{\eeq}{\end{equation}}
\newcommand{\bea}{\begin{eqnarray}}
\newcommand{\eea}{\end{eqnarray}}

\begin{document}


\title{Role of occupied $d$ bands in the dynamics of excited electrons and
holes in Ag}
\author{A. Garc\'\i a-Lekue$^{1}$, J. M. Pitarke$^{1,2}$, E. V.
Chulkov$^{2,3}$, A. Liebsch$^{4}$, and P. M. Echenique$^{2,3}$}
\affiliation{$^1$ Materia Kondentsatuaren Fisika Saila, Zientzi Fakultatea,
Euskal Herriko Unibertsitatea,\\
644 Posta kutxatila, E-48080 Bilbo, Basque Country\\
$^2$Donostia International Physics Center (DIPC) and Centro Mixto
CSIC-UPV/EHU,\\ Manuel de Lardizabal Pasealekua, E-20018 Donostia, Basque
Country\\
$^3$Materialen Fisika Saila, Kimika Fakultatea, Euskal Herriko
Unibertsitatea,\\ 
1072 Posta kutxatila, E-20080 Donostia, Basque Country\\
$^4$Institut f\"ur Festk\"orperforschung, Forschungszentrum J\"ulich,
52425 J\"ulich, Germany}                        

\date{\today}

\begin{abstract}
The role that occupied $d$ bands play in the inelastic lifetime of bulk and
surface states in Ag is investigated from the knowledge of the quasiparticle
self-energy. In the case of bulk electrons, $sp$ bands are taken to be
free-electron like. For surface states, the surface band structure of
$sp$ states is described with the use of a realistic one-dimensional
hamiltonian. The presence of occupied $d$ states is considered in both cases
by introducing a polarizable background. We obtain inelastic lifetimes of bulk
electrons that are in good agreement with first-principles band-structure
calculations. Our surface-state lifetime calculations indicate that the
agreement with measured lifetimes of both crystal-induced and image-potential
induced surface states on Ag(100) and Ag(111) is considerably improved when
the screening of $d$ electrons is taken into account.  
\end{abstract}

\pacs{71.45.Gm, 73.20.At, 73.50.Gr, 78.47.+p}

\maketitle

\section{Introduction}
Electron scattering in the bulk and at the surface
of solid materials has attracted  great interest, as
the Coulomb interaction between excited electrons and
the remaining electrons of the solid is fundamental
in many physical and chemical phenomena, such as charge transfer or
electronically induced adsorbate reactions at surfaces.
\cite{science1,science2,science3,nienhaus} A basic quantity in photoelectron
spectroscopy and quantitative surface analysis is the quasiparticle lifetime,
which sets the duration of the excitation and combined with the velocity 
yields the mean free path of the excited state.

Recently, two-photon photoemission (2PPE) and time-resolved 2PPE (TR-2PPE)
spectroscopies have provided accurate measurements of the lifetime of both bulk
and surface electrons with energies below the vacuum
level.\cite{expb,hofer1,Knoesel98,fauster2,fauster3,fauster1,new2,new1} 
The lifetime of bulk and surface holes below the Fermi level has been 
investigated by employing high-resolution angle-resolved photoemission (ARP)
techniques.\cite{theil,Hensberger,Valla,Straube,Gerlach,Reinert01,Balasubramanian,Tang}
The scanning tunneling microscope (STM) has also been used to determine the
lifetime of both surface-state holes and electrons in the noble metals Cu, Ag,
and Au.\cite{burgi,kliewer,Hovel,Yurov,Vitali,Pivetta} 

In the noble metals there are two kinds of surface states: intrinsic surface
states of $s$-$p_z$ symmetry (also called Shockley
states)\cite{shoc,gart,eber,himp} and image-potential induced surface
states.\cite{echenique,smith,rev} Shockley states are localized near the
topmost atomic layer and originate in the symmetry breaking at the surface.
Image states are Rydberg-like states trapped in the image-potential well
outside the surface of a solid with a projected band gap near the vacuum
level. Theoretical work based on many-body calculations of the electron
self-energy predicted both Shockley and image-state lifetimes that are, in the
case of Cu and other metal surfaces, in good agreement with
experiment.\cite{fauster2,fauster3,fauster1,Balasubramanian,
kliewer,imanol1,imanol2,chulkov0,german,fukui01,asier}  The key role
that occupied $d$ bands play in the dynamics of image-state electrons has also
been investigated,\cite{aran} showing that surface-plasmon decay yields a
surprisingly long lifetime of the first image state on the (100) surface of
Ag, in agreement with experiment.

In this paper, we report calculations of the inelastic lifetime of both bulk
and surface states in Ag, as obtained from the knowledge of the
complex quasiparticle self-energy by combining the dynamics of $sp$ valence
electrons with a simplified description of the occupied $d$ bands. In the case
of bulk electrons, $sp$ bands are taken to be free-electron like. For surface
states, single-particle wave functions and energies of $sp$ electrons are taken
to be the eigenfunctions and eigenenergies of a realistic one-dimensional
hamiltonian that describes the main features of the surface band structure.
The dynamically screened electron-electron (e-e) interaction is evaluated
within the random-phase approximation (RPA), and the GW approximation of
many-body theory is used to compute the complex self-energy. We present the
results of calculations of the lifetime of bulk excited electrons with
energies below the vacuum level, the first three image states on Ag(100), and
both the Shockley and first image state on Ag(111). We find that a realistic
description of the screening of $d$ electrons, which reduces the screened e-e
interaction and opens (in the case of image states) a new decay channel, is of
crucial importance for the understanding of the origin and magnitude of the
decay of both bulk and surface states.       

Unless stated otherwise, atomic units are used throughout, 
i.e., $e^2=\hbar=m_e=1$.       

\section{Theory}

In a system of many-electrons, the inverse quasiparticle lifetime is dictated
by the imaginary part of the complex energy of the quasiparticle. On the {\it
energy-shell} (i.e., neglecting the quasiparticle-energy renormalization), the
lifetime broadening of an excited state $\Phi({\bf r})$ with energy $E$ is
given by\cite{chem}  
\beq\label{500}
\tau^{-1}= -2\,\int\!d{\bf r}\int\!d{\bf r'}
\,\Phi^*({\bf r})\,{\rm Im}\,\Sigma({\bf r},{\bf r'};E)\,\Phi({\bf r'}).
\eeq
The complex self-energy $\Sigma({\bf r},{\bf r'};E)$ can be expanded in a
perturbative series of the so-called screened interaction $W({\bf r},{\bf
r'};E)$,\cite{fw} which describes the scattering of the  excited electron
with the remaining electrons of the system. In the GW approximation,
one considers only the first-order term in this expansion. If one further
replaces the exact one-electron Green function $G({\bf r},{\bf r'},E)$ by its
noninteracting counterpart $G^0({\bf r},{\bf r'},E)$, one finds
\beq\label{osma1.45}
{\rm Im}\Sigma({\bf r},{\bf r'};E)= 
\sum_{f}\Phi^*_f({\bf r'})\,{\rm Im}W({\bf r},{\bf r'};
E-E_f)\,\Phi_f({\bf r}).
\eeq
Here, the sum is extended over all available single-particle states
$\Phi_f({\bf r})$ with energy $E_f$ ($E_F\leq E_f\leq E$), $E_F$ is the Fermi
energy, and the screened interaction is
\begin{eqnarray}\label{72}
W({\bf r},{\bf r}';E)&=&v({\bf r},{\bf r'}) 
+\int\!d{\bf r}_1\int\!d{\bf r}_2\,v({\bf r},{\bf r}_1)\cr\cr
&\times&\chi({\bf r}_1,{\bf r}_2;E)\,v({\bf r}_2,{\bf r'}),
\end{eqnarray}
where $v({\bf r},{\bf r}')$ represents the bare Coulomb interaction and
$\chi({\bf r},{\bf r}';E)$ is the density-response function of the
many-electron system. In the random-phase approximation (RPA),
\begin{eqnarray}\label{chi0}
\chi({\bf r},{\bf r}',E)&=&\chi^0({\bf r},{\bf r'},E) 
+\int\!d{\bf r}_1\int\!d{\bf r}_2\,\chi^0({\bf r},{\bf r}_1,E)\cr\cr
&\times&v({\bf r}_1,{\bf r}_2)\,\chi({\bf r}_2,{\bf r'},E),
\end{eqnarray}
where $\chi^0({\bf r},{\bf r'},E)$ is the density-response function of
non-interacting electrons, which is obtained from a complete set of
single-particle wave functions and energies.

\subsection{Bulk states}

For a description of the electron dynamics of bulk states in Ag, we consider a
homogeneous assembly of interacting valence ($5s^1$) electrons immersed in a
polarizable background of $d$ electrons characterized by a local dielectric
function $\epsilon_d(\omega)$. Within this model, single-particle wave
functions are simply plane waves and the G$^0$W-RPA (also called G$^0$W$^0$)
lifetime broadening is obtained from Eqs. (\ref{500})-(\ref{chi0}) by
replacing the bare Coulomb interaction $v({\bf r},{\bf r}')$ by a modified
($d$-screened) Coulomb interaction $v'({\bf r},{\bf r}';\omega)$ of the form
\begin{equation}
v'({\bf r},{\bf r}';\omega)=v({\bf r},{\bf r}')\,\epsilon_d^{-1}(\omega).
\end{equation}
By introducing Fourier transforms, the lifetime broadening of an excited bulk
state of momentum ${\bf k}$ and energy $E=k^2/2$ is then found to be
\beq\label{eq1}
\tau_{{\bf k}}^{-1}=-{1\over 2\pi^2k}\int_0^{2k}
dq\,q\int_0^{\omega_{\rm max}}d\omega {\rm Im}W'_{q}(\omega),
\eeq
where $\omega_{\rm max}={\rm min}(E-E_F,kq-q^2/2)$,
and
\begin{equation}
W'_{q}(\omega)={v_{q}\over\epsilon_{q}(\omega)+\epsilon_d(\omega)-1}.
\end{equation}
Here, $v_{q}=4\pi/q^2$, $\epsilon_{q}(\omega)$ is the RPA
dielectric function of valence ($5s^1$) electrons
\begin{equation}
\epsilon_{q}(\omega)=1-v_{q}\,\chi^0_{q}(\omega),
\end{equation}
$\chi^0_{q}(\omega)$ being the Fourier transform of the density-response
function of non-interacting free electrons, i.e., the Lindhard
function.\cite{lind}

In the case of bulk states with energies very close to the Fermi level
($E-E_F<<E_F$) and in the limit of high electron densities, one finds
\begin{equation}\label{new}
\tau^{-1}_E={2(E-E_F)^2\over\pi q_F}\int_0^\infty{dq\over
q^4}\left[q_{TF}^2/q^2+\epsilon_d(\omega\to 0)\right]^{-2},
\end{equation}
where $q_{TF}$ is the Thomas-Fermi momentum [$q_{TF}=\sqrt{4q_F/\pi}$] and
$q_F$ is the Fermi momentum. The integral entering Eq. (\ref{new}) is easily
carried out to yield
\begin{equation}\label{last}
\tau^{-1}_{E}={(3\pi^2/2)^{1/3}\over
36}\,r_s^{5/2}(E-E_F)^2/\sqrt{\epsilon_d(\omega\to 0)},
\end{equation}
where $r_s$ is the electron-density parameter\cite{rs} of valence ($5s^1$)
electrons.

Eq. (\ref{new}) shows that the screening of $d$ electrons reduces the lifetime
broadening, this reduction being negligible in the long-wavelength ($q\to 0$)
limit where long-range interactions dominate. Eq.  (\ref{last}) shows that the overall
impact of $d$-electron screening is to decrease the lifetime broadening by a
factor of $\epsilon_d^{1/2}$, as suggested by Quinn.\cite{quinn}

\subsection{Surface states}

In the case of surface states, interacting valence ($5s^1$) electrons are
described by the eigenfunctions and eigenvalues of a realistic
one-dimensional hamiltonian that describes the main features of the surface
band structure,\cite{chulkov} and this bounded assembly of valence electrons is
then considered to be immersed in a polarizable background of $d$ electrons
which extends up to a certain plane $z=z_d$. Within this model there is
translational invariance in the plane of the surface, and the G$^0$W$^0$
lifetime broadening is obtained from Eqs. (\ref{500})-(\ref{chi0}), as in the
case of bulk states, by replacing the bare Coulomb interaction $v({\bf r},{\bf
r}')$ by a modified (d-screened) Coulomb interaction $v'({\bf r},{\bf
r}';\omega)$ whose two-dimensional (2D) Fourier transform yields (the $z$ axis
is taken to be perpendicular from the surface)\cite{catalina}
\begin{eqnarray}\label{ll} v'_{{\bf q}_\parallel}(z,z';\omega)&=&\frac{2\pi}
{q_\parallel\,\epsilon_d(z',\omega)}\, [{\rm e}^{-q_\parallel\,|z-z'|}+{\rm
sgn}(z_d-z')\nonumber\\ &\times&\sigma_d(\omega)\,{\rm
e}^{-q_\parallel|z-z_d|}{\rm e}^{-q_\parallel|z_d-z'|}], \end{eqnarray} where
\begin{equation}\label{epsd}
\epsilon_d(z,\omega)=\cases{\epsilon_d(\omega),&$z\leq z_d$ \cr\cr 1, & $z >
z_d$} \end{equation} and \begin{equation}
\sigma_d={\epsilon_d(\omega)-1\over\epsilon_d(\omega)+1}.
\end{equation}
The first term in Eq. (\ref{ll}) is simply the bare Coulomb interaction
$v_{{\bf q}_\parallel}(z,z')=2\pi\,{\rm e}^{-q_\parallel|z-z'|}/q_\parallel$
screened by the polarization charges induced within the polarizable
background. The second term stems from polarization charges at the boundary of
the medium. 

By introducing 2D Fourier transforms, one finds the following expression for
the lifetime broadening of an excited state ${\rm e}^{{\rm i}{\bf
k}_\parallel\cdot{\bf r}_\parallel}\phi(z)$ of energy
$E=k_\parallel^2/(2m)+\varepsilon$:
\begin{equation}\label{taum1}
\tau_{{\bf k}_\parallel,\varepsilon}^{-1}=-2\,\int\!\!dz\int\!\!
dz'\,\phi^{*}(z)\,{\rm Im}\Sigma_{{\bf k}_\parallel,\varepsilon}(z,z')\,\phi(z'),
\end{equation}
where
\begin{equation}\label{selfener}
{\rm Im}\Sigma_{{\bf k}_\parallel,\varepsilon}(z,z')=
\sum_{{\bf q}_\parallel,\varepsilon_f}\phi^{*}_f(z')\,{\rm
Im}W'_{{\bf q}_\parallel}(z,z';E-E_f)\,\phi_f(z),
\end{equation}
the sum being extended over all available single-particle states
${\rm e}^{{\rm i}({\bf
k}_\parallel-{\bf q}_\parallel)\cdot{\bf r}_\parallel}\phi_f(z)$ of energy
$E_f=({\bf k}_\parallel-{\bf q}_\parallel)^2/(2m_f)+\varepsilon_f$ ($E_F\leq
E_f\leq E$). The screened interaction is
\begin{eqnarray}\label{72p}
W_{{\bf q}_\parallel}'(z,z';\omega)&=&v'_{{\bf q}_\parallel}(z,z';\omega) 
+\int\!dz_1\int\!dz_2\,v'_{{\bf q}_\parallel}(z,z_1;\omega)\cr\cr
&\times&\chi_{{\bf q}_\parallel}(z_1,z_2';\omega)\,v'_{{\bf
q}_\parallel}(z_2,z';\omega), 
\end{eqnarray}
where
\begin{eqnarray}\label{chi1}
\chi_{{\bf q}_\parallel}(z,z';\omega)&=&\chi^0_{{\bf q}_\parallel}(z,z';\omega)
+\int\!dz_1\int\!dz_2\,\chi^0_{{\bf
q}_\parallel}(z,z_1;\omega)\cr\cr
&\times&v'_{{\bf q}_\parallel}(z_1,z_2;\omega)\,\chi_{{\bf
q}_\parallel}(z_2,z';\omega),
\end{eqnarray} and
\begin{eqnarray}\label{chi2}
\chi^0_{{\bf q}_\parallel}(z,z';\omega)&=&{2\over A}\,\sum_{{\bf
k}_\parallel;\varepsilon_i,\varepsilon_j}\phi_i(z)\phi_j^*(z)\phi_j(z')\phi_i^*(z')\cr\cr
&\times&{\theta(E_F-E_i)-\theta(E_F-E_j)
\over E_i-E_j+(\omega+{\rm i}\eta)},
\end{eqnarray}
with $A$ being the normalization area, $E_i=\varepsilon_i+k_\parallel^2/2$,
$E_j=\varepsilon_j+({\bf k}_\parallel+{\bf q}_\parallel)^2/2$, $\theta(E)$ the
Heaviside function, and $\eta$ a positive infinitesimal.

We take all the single-particle wave functions and
energies entering Eqs. (\ref{taum1})-(\ref{chi2}) to be the eigenfunctions and
eigenvalues of the one-dimensional hamiltonian of Ref. \onlinecite{chulkov},
and we account for the potential variation parallel to the surface through the
introduction of the effective masses $m$ and $m_f$ entering the expressions
for the total energies $E$ and $E_f$.

Finally, we note that in the $z_d\to -\infty$ limit
[or, equivalently, if $\epsilon_d(z,\omega)$ is taken to be equal to unity for
all $z$], $v'_{{\bf q}_\parallel}(z,z';\omega)$ coincides with the 2D Fourier
transform of the bare Coulomb interaction and our lifetime broadening
$\tau_{{\bf k}_\parallel,\varepsilon}$ reduces to that of excited states in
the absence of $d$ electrons, as calculated in previous
work.\cite{imanol1,imanol2}

\section{Results and discussion}

\subsection{Bulk states}

Silver is a noble metal with $5s^1$ valence electrons and entirely filled
$4d$-like bands. Valence electrons form an electron gas with $r_s=3.02$.
Assuming that the only effect of $d$ electrons is to modify the electron
screening of the solid, these electrons are expected to form a polarizable
background characterized by a local dielectric function $\epsilon_d(\omega)$
which we take from bulk optical data.\cite{johnson} In the static limit
($\omega\to 0$), $\epsilon_d\sim 4$; in the absence of $d$ electrons,
$\epsilon_d(\omega)$ would simply be equal to unity.

\begin{figure}[hbt!]
\includegraphics[width=0.45\textwidth,height=0.3375\textwidth]{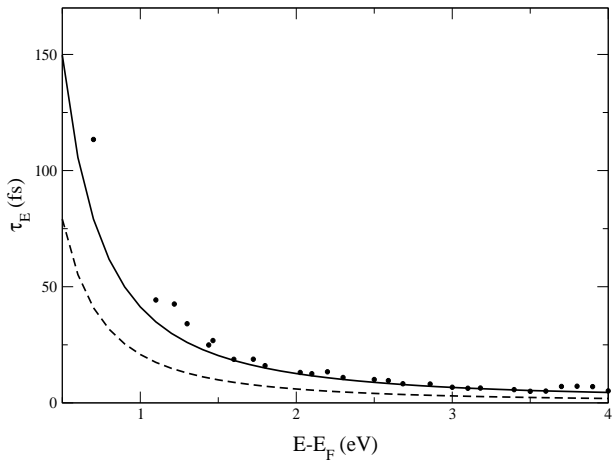}
\caption{ Lifetime of bulk excited electrons  in Ag versus the electron energy
(with respect to the Fermi level), as obtained from Eq. (\ref{eq1}) both in the
presence (solid line) and in the absence (dashed line) of $d$ electrons. Solid
circles represent the result of {\it on-the-energy-shell} first-principles
calculations reported in Ref.~\onlinecite{idoia}.}
\label{fig1} \end{figure}

Fig.~\ref{fig1} shows the lifetime of bulk excited electrons in Ag versus the
electron energy (with respect to the Fermi level), as obtained from Eq.
(\ref{eq1}) both in the presence (solid line) and in the absence (dashed
line) of $d$ electrons. $d$ electrons give rise to additional screening, thus
increasing the lifetime of all electrons above the Fermi level. We also compare
our results (solid line) with first-principles calculations of the inelastic
lifetime of bulk excited electrons in Ag\cite{idoia} (solid circles), showing
a remarkable agreement for all electron energies under consideration. This
indicates that, for excited states within a few eV above the Fermi level, our
simplified description of the $d$ bands in Ag [which dominate the density of
states (DOS) with energies from $\sim 4\,{\rm eV}$ below the Fermi level] in
terms of a polarizable medium is adequate.

\vspace{0.75cm}

In order to analyse the impact of $d$-electron screening as a function of the
momentum transfer, we define the lifetime broadening $\tau^{-1}_{\bf k}(q_{\rm
max})$ by replacing the actual maximum momentum transfer $2k$ entering Eq.
(\ref{eq1}) by $q_{\rm max}$ [which we vary from zero to its
actual value $2k$], and consider the ratio
\begin{equation}\label{ratio}
r(q_{\rm max})={\int_0^{2q_{\rm max}}
dq\,q\int_0^{\omega_{\rm max}}v_q\,{\rm Im}\epsilon^{-1}_q(\omega)
\over \int_0^{2q_{\rm max}}
dq\,q\int_0^{\omega_{\rm max}}{\rm Im}W'_{q}(\omega)}.
\eeq
We have plotted this ratio in Fig. ~\ref{fig2} versus $q_{\rm max}$ for bulk
states with energies very near the Fermi level,
showing that in the long-wavelength ($q\to 0$) limit $d$ electrons do not
participate in the screening of e-e interactions [see also Eq. (\ref{new})].
Hence, if the electron or hole decay were dominated by processes in which the
momentum transfer is small, as occurs in the case of Shockley surface states
in Ag(111), the overall impact of the screening of $d$ electrons should be
expected to be small. 

\vspace{0.75cm}
\begin{figure}[hbt!]
\vspace{0.75cm}
\includegraphics[width=0.45\textwidth,height=0.3375\textwidth]{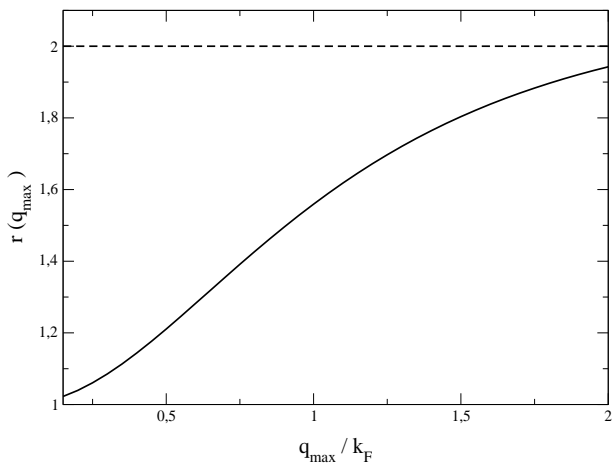}
\caption{Ratio $r(q_{\rm max})$ between the lifetime broadenings
$\tau^{-1}_{\bf k}(q_{\rm max})$ in the abscence [numerator of Eq.
(\ref{ratio})] and in the presence [denominator of Eq. (\ref{ratio})] of $d$
bands, as a function of $q_{\rm max}$ and for bulk states with energies very
near the Fermi level ($E-E_F<<1$). For these energies, the ratio of Eq.
(\ref{ratio}), which does not depend on the electron energy $E$, approaches 
at $q_{\rm max}=2k_F$ the result one would obtain from Eq.
(\ref{last}), i.e., $r(2k)\sim\sqrt{\epsilon_d(\omega\to
0)}\sim 2$.}
\label{fig2} \end{figure}

\subsection{Surface states}

The projected band gaps and available bulk states on the (100) and (111)
surfaces of Ag are shown in Fig.~\ref{fig3} at the $\Gamma$ point
($k_\parallel=0$). In the case of Ag(100), the surface electronic structure
supports the whole Rydberg series of image states lying close to the center of
the projected band gap. The first three image states have binding energies of
0.53, 0.17, and 0.081 eV ($\varepsilon-E=3.90, 4.26$, and $4.35\,{\rm eV}$),
and their probability-densities have maxima at 3.8, 14.5, and $32.6\,\AA$
outside the crystal edge ($z=0$), which we choose to be located half a lattice
spacing beyond the last atomic layer. In the case of Ag(111), the surface
electronic structure supports the first ($n=1$) image state with a binding
energy of 0.77 eV ($\varepsilon-E_F=3.79$), the probability-density being
maximum at $0.1\,\AA$ inside the crystal edge, and a partially occupied
Shockley ($n=0$) surface state with energy lying at 0.065 eV below the Fermi
level and probability maximum at $2.2\,{\rm\AA}$ outside the crystal edge.

\vspace{00.75cm}
\begin{figure}[hbt!]
\includegraphics[width=0.5\textwidth]{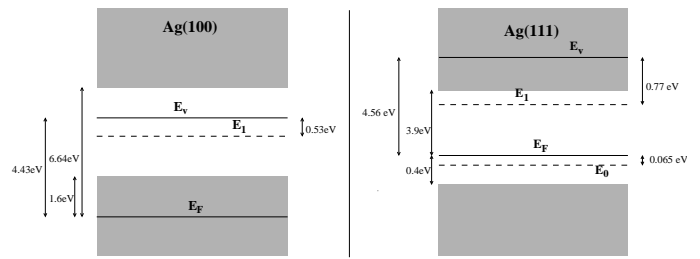}
\caption{Projected band gaps and available bulk states on the (100) and (111)
surfaces of Ag for $k_\parallel=0$. The shaded area represents the projected
bulk bands.}
\label{fig3}
\end{figure}

The effective mass of image states is known to be close to the free-electron
mass ($m\sim 1$), whereas the effective mass of the $n=0$ surface state on
Ag(111) is found to be $m\sim 0.44$. For all single-particle states below the
bottom of the band gap we choose the effective mass to increase from our
computed values of 0.45 [Ag(100)] and 0.42 [Ag(111)] at the bottom of the gap
to the free-electron mass at the bottom of the valence band. The boundary of
the polarizable medium is taken to be at $z_d=-1.5\,a_0$ inside the crystal,
which was previously found to best reproduce the anomalous dispersion
of surface plasmons in Ag.\cite{liebsch}

\subsubsection{Surface-plasmon excitation}

The lifetime broadening of excited states of energy $E$ originates in processes
in which the excited electron/hole decays into an empty state above/below the
Fermi level. These processes can be realized by transferring energy and
momentum to an excitation of the medium, thereby creating either an
electron-hole pair or a collective excitation of energy $\omega<|E-E_F|$.
Without $d$-electron screening, the Ag surface-plasmon energy
($\omega_{sp}=6.5\,{\rm eV}$) is larger than the excitation energies
($|E-E_F|$) under consideration, and only electron-hole pairs can be created.
However, the presence of $d$ electrons is known to reduce this surface-plasmon
energy from $\sim 6.5$ to $\sim 3.7\,{\rm eV}$,\cite{liebsch} so that
relaxation of image states via surface-plasmon excitation becomes feasible and
contributes to the lifetime broadening of Eq. (\ref{taum1}).

\vspace{0.75cm}
\begin{figure}[!hbt]
\includegraphics[width=0.45\textwidth]{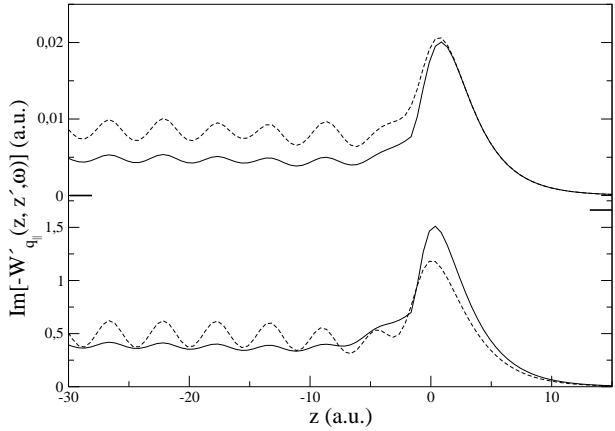}
\caption{Imaginary part of the screened interaction ${\rm Im}\left[-W'_{{\bf
q}_\parallel}(z,z';\omega)\right]$ versus $z$ for $z'=z$, as obtained from Eq.
(\ref{72p}) in the presence (solid lines) and in the abscense (dashed lines) of
$d$ electrons. The parallel momentum transfer is taken to be
$q_\parallel=0.3\,q_F$ and two different values are considered for the
energy transfer: $\omega=0.05\,{\rm eV}$ (top panel) and $\omega=4\,{\rm
eV}$ (bottom panel). }
\label{fig4}
\end{figure}

A key ingredient in the decay mechanism of excited states is the imaginary
part of the screened interaction ${\rm Im}\left[-W'_{{\bf
q}_\parallel}(z,z';\omega)\right]$ entering Eq. (\ref{selfener}).
Fig.~\ref{fig4} shows this quantity versus the $z$ coordinate both in the
presence (solid lines) and in the absence (dashed lines) of $d$ electrons, for
$z'=z$, $q_\parallel=0.3\,q_F$, and two choices of the energy transfer:
$\omega=0.05$ and  $4\,{\rm eV}$. $d$ electrons give rise to additional
screening; hence, in the interior of the solid they simply reduce ${\rm
Im}W'$, as in the case of bulk states. However, when the energy transfer
approaches the reduced surface-plasmon energy of $\sim 3.7\,{\rm eV}$ (see
bottom panel of Fig.~\ref{fig4}), this effect is outweighed near the surface
by the opening of the surface-plasmon excitation channel which only occurs in
the presence of $d$ electrons (solid lines). Hence, while for
$\omega=0.05\,{\rm eV}$ the presence of $d$ electrons reduces ${\rm
Im}\left[-W'\right]$ both inside the solid and near the surface, for
$\omega=4\,{\rm eV}$ $d$-electron screening enhances ${\rm Im}\left[W'\right]$
near the surface.

\vspace{0.75cm}
\begin{figure}[hbt!]
\includegraphics[width=0.45\textwidth]{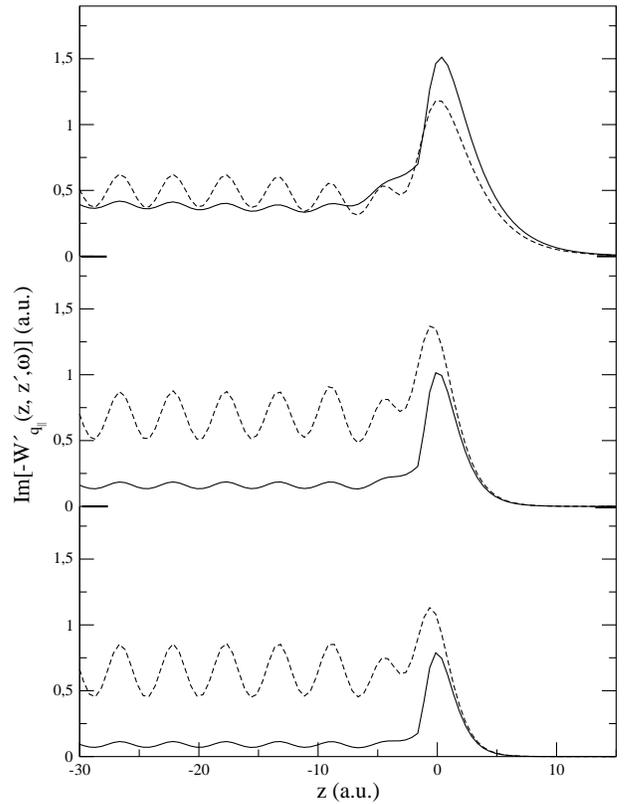}
\vspace{0.3cm}
\caption{Imaginary part of the screened interaction ${\rm Im}\left[-W'_{{\bf
q}_\parallel}(z,z';\omega)\right]$ versus $z$ for $z'=z$, as obtained from Eq.
(\ref{72p}) in the presence (solid lines) and in the abscense (dashed lines) of
$d$ electrons. Three different values are considered for the parallel momentum
transfer: $q_\parallel=0.3\,q_F$ (top panel), $q_\parallel=0.6\,q_F$ (middle
panel), and $q_\parallel=0.8\,q_F$ (bottom panel). The energy transfer is
taken to be $\omega=4\,{\rm eV}$. }
\label{fig5}
\end{figure} 

Fig.~\ref{fig5} shows the imaginary part of the screened interaction versus
the $z$ coordinate, as in Fig.~\ref{fig4}, but now for $\omega=4\,{\rm eV}$
and three different values of the momentum transfer: $q_\parallel=$ 0.3, 0.6,
and 0.8 $q_F$. For the lowest value of $q_\parallel$ (see also
Fig.~\ref{fig4}) surface plasmons can be excited in the presence of $d$
electrons, which yields an enhancement of ${\rm Im}\left[-W'\right]$ near the
surface. As $q_\parallel$ increases the impact of $d$-electron screening is
more pronounced in the interior of the solid, as in the case of bulk states,
but there is no enhacement of ${\rm Im}\left[-W'\right]$ near the surface,
showing that surface-plasmon excitation does not occur for large values of the
parallel momentum transfer.

\subsubsection{Image states on Ag(100)}

Now we consider the decay of image states on Ag(100) at the $\bar\Gamma$
point. At this point, the energy of the $n=1$ image state lies 3.9
eV above the Fermi level. Hence, the decay of this excited state can be
realized by creating excitations of energy $\omega\leq 3.9\,{\rm eV}$, which
in the presence of $d$ electrons can be either electron-hole pairs or surface
plasmons.

Calculations of the imaginary part of the Ag(100) $n=1$ image-state
self-energy ${\rm Im}\left[-\Sigma_{{\bf
k}_\parallel,\varepsilon}(z,z')\right]$ entering Eq. (\ref{taum1}) were
reported in Ref.~\onlinecite{aran}, showing that, as in the case of
${\rm Im}\left[-W'\right]$ (see Figs.~\ref{fig4} and \ref{fig5}), the reduction
of the self-energy inside the solid due to the additional screening of $d$
electrons is outweighed near the surface by the opening of the surface-plasmon
excitation channel.

\vspace{0.75cm}
\begin{figure}[hbt!]
\includegraphics[width=0.45\textwidth,height=0.3375\textwidth]{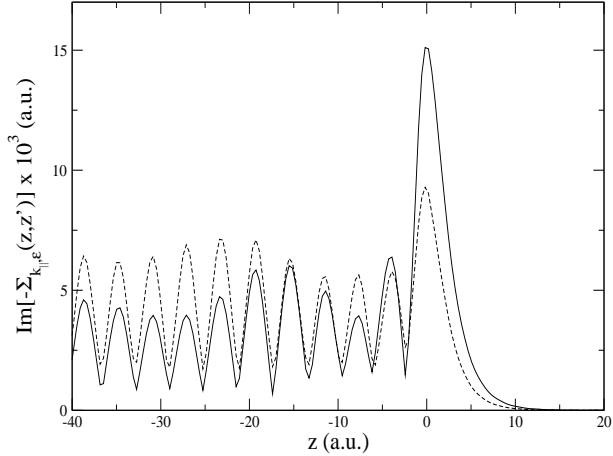}
\caption{Maximum of the imaginary part of the $n=1$ image-state self-energy 
${\rm Im}\left[-\Sigma_{{\bf k}_\parallel,\varepsilon}(z,z')\right]$, versus
$z$, in the vicinity of the (100) surface of Ag, as obtained from Eq.
(\ref{selfener}) both on the presence (solid lines) and in the abscense
(dashed lines) of $d$ electrons.  For each value of $z$, ${\rm
Im}\left[-\Sigma\right]$ is evaluated at the $z'$ coordinate for which it is
maximum. Inside the solid this occurs at $z'=z$, as in the case of bulk
excited states; nevertheless, for $z$ coordinates in the vacuum side of the
solid the maximum of ${\rm Im}\left[-\Sigma\right]$ lags behind and remains
localized at $z\sim 0$ rather than $z'=z$, showing a highly nonlocal behaviour
in the presence of a metal surface.} \label{fig6}
\end{figure} 

The magnitude of the maximum of the imaginary part of the Ag(100)
$n=1$ image-state self-energy is plotted in Fig.~\ref{fig6}, as a function of
$z$. This is an oscillating function of $z'$ within the bulk and
reaches its highest value near the surface. The oscillatory behaviour within
the bulk is dictated by the periodicity of the final-state wave functions
$\phi_f(z)$ entering Eq. (\ref{selfener}), and the highest value near the
crystal edge is the result of the fact that electron-hole pair creation is
enhanced in the vicinity of the surface. We note that the presence of $d$
electrons reduces the maximum of ${\rm Im}\left[-\Sigma\right]$ in the interior
of the solid but enhances this quantity near the surface, due to the opening of
the surface-plasmon excitation channel. Hence, one might expect from Eq.
(\ref{taum1}) that this enhancement of ${\rm Im}\left[-\Sigma\right]$ near the
surface should yield an accordingly larger lifetime broadening. The key point
of Ref.~\onlinecite{aran} was, however, to show that this trend is reversed,
due to the characteristic non-locality of the self-energy near the surface.

\vspace{0.75 cm}
\begin{figure}[!hbt]
\includegraphics[width=0.45\textwidth,height=0.3375\textwidth]{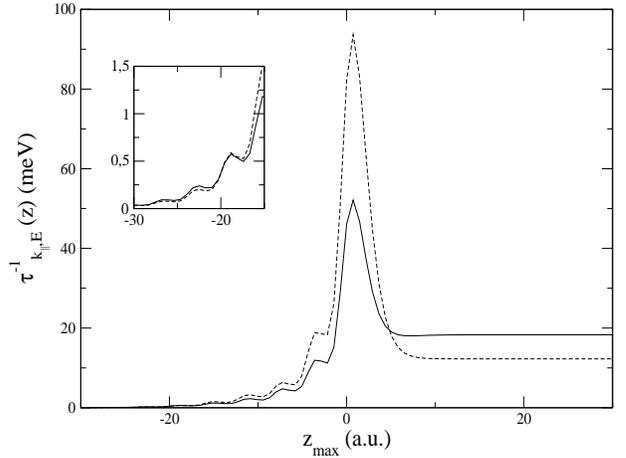}
\caption{Lifetime broadening $\tau^{-1}_{{\bf k}_\parallel,E}(z_{\rm max})$ of
the $n=1$ image state at the $\bar\Gamma$ point of Ag(100), as obtained from
Eq. (\ref{taum1}) by replacing the upper limit ($z\to\infty$) of the integral
over $z$ by $z_{\rm max}$. Solid and dashed lines represent the result
obtained in the presence and in the abscence of $d$ electrons, respectively.}
\label{fig7}
\end{figure} 

In order to understand this result, we define the lifetime broadening 
$\tau^{-1}_{{\bf k}_\parallel,\varepsilon}(z_{\rm max})$ by replacing the upper
limit ($z\to\infty$) entering the first integral of Eq. (\ref{taum1}) by
$z_{\rm max}$ [which we vary from $-\infty$ to its actual value $z_{\rm
max}\to\infty$]. We have plotted in Fig.~\ref{fig7} the lifetime broadening
$\tau^{-1}_{{\bf k}_\parallel,\varepsilon}(z_{\rm max})$ of the $n=1$ image
state on Ag(100). For $z_{\rm max}$ well inside the
solid (see inset of Fig.~\ref{fig7}), coupling of the image state with the
crystal only occurs through the penetration of the image-state wave function
into the solid, where the presence of $d$ electrons reduce the decay rate. As
$z_{\rm max}$ approaches the surface, coupling of the image state with the
crystal is still dominated by the penetration of the image-state wave function
into the solid, but now in the presence of $d$ electrons the opening of the
surface-plasmon decay channel increases the decay rate. Finally, as $z_{\rm
max}$ moves into the vacuum the imaginary part of the self-energy exhibits a
highly non-local behaviour (for $z$ outside the solid the main peak of ${\rm
Im}[-\Sigma]$ lags behind and remains localized at $z\sim 0$), so that
coupling mainly occurs through interference between the image-state amplitude
at $z>0$ and the image-state amplitude itself at $z'\sim 0$. Since this
amplitude has a node at $z\sim a_0$ (see Ref.~\onlinecite{aran}) and is
negative at $z'\sim 0$ where ${\rm Im}\left[-\Sigma\right]$ is maximum,
interference contributions to the lifetime broadening must be negative. As a
result, the lifetime broadening $\tau^{-1}_{{\bf
k}_\parallel,\varepsilon}(z_{\rm max})$ is reduced when $z_{\rm max}$ crosses
the surface. Since the presence of $d$ electrons enhances the self-energy near
the surface, negative interference is also enhanced and Fig.~\ref{fig7} shows
that the net effect of decay via surface plasmons is a considerably reduced
lifetime broadening as $z_{\rm max}\to\infty$.

\begin{table}[hbtp!]
\caption{Calculated lifetime broadening $\hbar\tau^{-1}$ (in meV) of the
$n=1,2$, and 3 image states on Ag(100) either in the presence of a polarizable
background of $d$ electrons which extends up to a plane located at
$z_d=-1.5\,a_0$ or in the absence of $d$ electrons ($z_d\to -\infty$). Local
and nonlocal contributions to the lifetime broadening have been obtained by
confining the integrals in Eq. (\ref{taum1}) to $z,z'<0$ (local) and by
confining the these integrals to either $z>0$ or $z'>0$ (nonlocal). The
experimental linewidth is taken from TR-2PPE experiments.\cite{hofer1} The
lifetime in fs ($1\,{\rm fs}=10^{-15}\,s$) is obtained by noting that
$\hbar=658\,{\rm meV}\,{\rm fs}$.}
\begin{tabular}{lcccccc} \hline
 n&\,\,\,\,\,\,&$z_d$& Local  & Nonlocal  &
 Total & Experiment \\ \hline\hline
1&& -$\infty$ & 34 & -16  &18&  \\ 
 && -$1.5$ & 59 & -47  &12& 12\\ \cline{1-7}
2&& -$\infty$ &  9&-5   &4&  \\ 
 && -$1.5$ &14  &-11   &3&4\\ \cline{1-7}
3&& -$\infty$ &5  &-3   &2&  \\ 
 && -$1.5$ & 6 &-5   &1& 1.8\\ \hline
\end{tabular}
\label{table1}
\end{table}

We have calculated from Eq. (\ref{taum1}) separate contributions to the decay
of the first three image states on Ag(100) arising from $z,z'<0$ ("local"
part) and from $z>0$ or $z'>0$ ("nonlocal" part), and we have obtained the
results presented in Table~\ref{table1}. The TR-2PPE measurements reported in
Ref.~\onlinecite{hofer1} are also presented in this table. Our calculations
indicate that the impact of $d$-electron screening on the negative
interference near the surface outweighs the increase of the local part of the
decay, which yields a reduced lifetime broadening in agreement with experiment.

\subsubsection{ $n=0$ and $n=1$ surface states on Ag(111)}

Fig.~\ref{fig8} shows the dispersion of the $n=0$ Shockley state on the (111)
surface of Ag. At the $\bar\Gamma$ point this is an occupied state; hence, a
hole can be created at this point, which will decay through the coupling with
bulk states (interband transitions) and through the coupling, within the
surface state itself, with surface states of different wave vector parallel to
the surface (intraband transitions). Previous calculations of the lifetime
broadening of this Shockley hole showed that the broadening arising  from
intraband transitions represents $\sim 90\%$ of the total decay rate.
\cite{kliewer,german} As the momentum transfer involved in these transitions
(see Fig.~\ref{fig8}) is very small, the presence of a polarizable medium of
$d$ electrons is not expected to play an important role.

\vspace{0.75cm}
\begin{figure}[hbt!]
\vspace{0.75cm}
\includegraphics[width=0.45\textwidth,height=0.3375\textwidth]{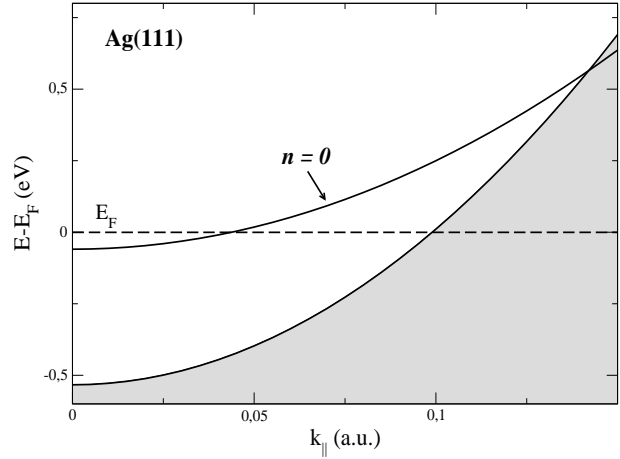}
\caption{The solid line represents the dispersion of the Shockley ($n=0$)
surface state on Ag(111). The shaded area represents the projected bulk bands
in this surface.}
\label{fig8}
\end{figure}  

\vspace{0.7cm}
\begin{figure}[hbt!]
\vspace{0.7cm}
\includegraphics[width=0.45\textwidth]{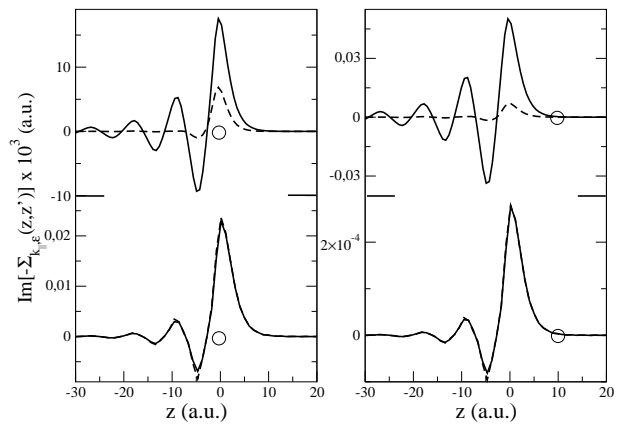}
\vspace{0.3cm}
\caption{Imaginary part of the $n=0$ (bottom panels) and $n=1$ (top panels)
surface-state self-energy ${\rm Im}\left[-\Sigma_{{\bf
k}_\parallel,\varepsilon}(z,z')\right]$ versus $z$ in the
vicinity of the (111) surface of Ag, both in the presence (solid lines) and in
the absence (dashed lines) of $d$ electrons. The value of the $z'$ coordinate
(indicated by an open circle) is fixed at $z'=0$ (left panels) and
$z=10\,a_0$ (right panels). The geometrical electronic edge ($z=0$ is chosen
to be located half an interlayer spacing beyond the last atomic layer.
Parallel momentum is taken to be ${\bf k}_\parallel=0$. $n=0$ and $n=1$
surface-state energies are $\varepsilon=E_F+3.79\,{\rm eV}$ and
$\varepsilon=E_F-0.065\,{\rm eV}$, respectively.}
\label{fig9}
\end{figure}  

The bottom panel of Fig.~\ref{fig9} exhibits the imaginary part of the
$n=0$ Shockley-hole self-energy at the $\bar\Gamma$ point both in the presence
(solid lines) and in the absence (dashed lines) of $d$ electrons, as a
function of $z$ for fixed values of the $z'$ coordinate: $z'=0$ and $10\,a_0$.
This figure clearly shows that the impact of $d$-electron screening on the
decay of this surface state is nearly negligible. In contrast, the screening
of $d$ electrons opens a surface-plasmon decay mechanism for the $n=1$ image
state on Ag(111) (see the top panel of Fig.~\ref{fig9}), as occurs in the case
of Ag(100).

In Table~\ref{table2} we summarize the results we have obtained for the
lifetime broadening of both the $n=0$ and $n=1$ surface states on Ag(111) at
the $\bar\Gamma$ point. We observe that the broadening is always reduced when
$d$-electron screening participates. The lifetime broadening of the Shockley
hole ($n=0$) is only slightly reduced, because of the weak screening of $d$
electrons involved in the interaction with the Fermi sea. In the case of the
$n=1$ image state, however, both local ($z<0$ and $z'<0$) and nonlocal ($z>0$
or $z'>0$) contributions to the broadening are enhanced in the presence of
$d$-electron screening, due to the opening of the surface-plasmon decay
channel. Nevertheless, the role that $d$ electrons play enhancing the nonlocal
contribution to the decay, which is dominated by negative interference, is
more pronounced than the role that these electrons play enhancing the local
decay. As a result, the net effect of decay via surface plasmons is a
considerably reduced lifetime broadening, as occurs in the case of image
states on Ag(100).

\begin{table}[hbtp!]
\caption{Calculated lifetime broadening $\hbar\tau^{-1}$ (in meV) of the
Shockley ($n=0$) and image ($n=1$) surface states on Ag(111) either in the
presence of a polarizable background of $d$ electrons which extends up to a
plane located at $z_d=-1.5\,a_0$ or in the absence of $d$ electrons ($z_d\to
-\infty$). Local and nonlocal contributions to the lifetime broadening have
been obtained by confining the integrals in Eq. (\ref{taum1}) to $z,z'<0$
(local) and by confining the these integrals to either $z>0$ or $z'>0$
(nonlocal). The experimental linewidths of Shockley and image states are taken
from Refs.~\onlinecite{kliewer} and~\onlinecite{hofer1}, respectively. The
lifetime in fs ($1\,{\rm fs}=10^{-15}\,s$) is obtained by noting that
$\hbar=658\,{\rm meV}\,{\rm fs}$.}
\begin{tabular}{lcccccc} \hline
 n&\,\,\,\,\,\,&$z_d$& Local  & Nonlocal  &
 Total & Experiment \\ \hline\hline
0&& -$\infty$ &1  &2   &3&  \\ 
 && -$1.5$ & 1 & 1.8  &2.8&6\\ \cline{1-7}
1&& -$\infty$ &48  &-4   &44&  \\ 
 && -$1.5$ & 103 & -67  &36& 22\\ \hline
\end{tabular}
\label{table2}
\end{table}

Also shown in Table~\ref{table2} are low-temperature STM
measurements\cite{kliewer} of the lifetime of an excited hole at the edge
($\bar\Gamma$ point) of the partially occupied $n=0$ surface
state and the 2PPE measurements reported by McNeil {\it et al.}
for the $n=1$ image state,\cite{expb} both on the (111) surface of
Ag. At this point, we note that electron-phonon scattering is
expected to yield a contribution to the $n=0$ surface-state linewidth of
$3.9\,{\rm meV}$,\cite{asier} which would result in a total linewidth of
$6.7\,{\rm meV}$ in good agreement with the experimentally measured linewidth
of $6\,{\rm meV}$. On the other hand, our calculations show that the inclusion
of the surface-plasmon decay channel leads to a better agreement of the
calculated $n=1$ image-state linewidth with experiment, although our
calculated linewidth for this surface state is still too high. 

\section{Summary and conclusions}

We have investigated the role that occupied $d$ bands play in the lifetime of
bulk and surface states in Ag, by combining an accurate description of the
dynamics of $sp$ valence electrons with a physically motivated model in which
the occupied $d$ bands are accounted for by the presence of a polarizable
medium. In the case of bulk excited states, we have obtained lifetimes that are
in good agreement with first-principles band-structure calculations. Our
surface-state lifetime calculations indicate that the agreement with measured
lifetimes of both the $n=0$ crystal-induced and the lowest-lying $n=1$
image-potential induced surface states on silver surfaces is considerably
improved when the screening of $d$ electrons is taken into account.

The impact of $d$-electron screening on the dynamics of the Shockley surface
hole on Ag(111) is entirely due to the reduction of the screened
interaction between the excited hole and the Fermi sea. Since the decay of the
excited hole is dominated by intraband transitions within the Shockley surface
band itself, where the parallel momentum transfer is always very small, the
screening of $d$ electrons is comparatively weak and the lifetime broadening is
reduced by no more than $10\%$.

As far as image states are concerned, the reduction (in the presence of $d$
electrons) of the surface-plasmon energy from 6.5 to 3.7 eV opens a new decay
channel that enhances the image-state self-energy and more than compensates the
reduction of the screened e-e interaction. Nevertheless, our results
demonstrate that the highly nonlocal character of the self-energy near the
surface ultimately leads to a reduced lifetime broadening of image states on
both the (100) and (111) surfaces of Ag. In the case of the $n=1$ image
state on Ag(100), the reduced lifetime broadening is in excellent agreement
with the experiment. The agreement with the  measured lifetime of the $n=1$
image state on Ag(111) is also improved when the screening of $d$ electrons is
considered, although in this case the experimental linewidth is still well
above the theoretical prediction. The experimental linewidths of the $n=2$ and
$n=3$ image states on Ag(100) are also slightly above our best theoretical
prediction, as occurs in the case of the corresponding image states on
Cu(100).\cite{imanol2} As the linewidth of excited image states ($n>1$) is only
of a few meV's, small discrepancies with the experiment may be attributed to a
combination of scattering with phonons, many-body effects beyond the $GW$
approximation, and the necessity of an {\it ab initio} description of the
dynamical response of both $sp$ and $d$ electrons in the noble metals. 

\acknowledgements

We acknowledge partial support by the University of the Basque Country, the
Basque Unibertsitate eta Ikerketa Saila, the Spanish Ministerio de Ciencia y
Tecnolog\'\i a, and the Max Planck Research Award Funds.


\begin{thebibliography}{100}
\bibitem{rs} $r_s$ is defined as $(4\pi/3)(r_sa_0)^3=1/n$, $n$
being the electron density and $a_0$ the Bohr radius [$a_0=0.529\,\AA$].
\end{thebibliography}
\end{document}